\documentclass{bmvc2k}
\usepackage{wrapfig}

\usepackage{pifont}

\newcommand{\cmark}{\ding{51}} % ✓
\newcommand{\xmark}{\ding{55}}
\title{PanoHair: Detailed Hair Strand Synthesis on Volumetric Heads}

\addauthor{Shashikant Verma}{shashikant.verma@iitgn.ac.in}{1}
\addauthor{Shanmuganathan Raman}{shanmuga@iitgn.ac.in}{1}

\addinstitution{
 CVIG Lab \\
 Indian Institute of Technology Gandhinagar\\
 India
}

\runninghead{Verma et. al. }{PanoHair}

\begin{document}

\maketitle

\begin{abstract}
Achieving realistic hair strand synthesis is essential for creating lifelike digital humans, but producing high-fidelity hair strand geometry remains a significant challenge. 
Existing methods require a complex setup for data acquisition, involving multi-view images captured in constrained studio environments. Additionally, these methods have longer hair volume estimation and strand synthesis times, which hinder efficiency.
We introduce PanoHair, a model that estimates head geometry as signed distance fields using knowledge distillation from a pre-trained generative teacher model for head synthesis. 
Our approach enables the prediction of semantic segmentation masks and 3D orientations specifically for the hair region of the estimated geometry. 
Our method is generative and can generate diverse hairstyles with latent space manipulations. For real images, our approach involves an inversion process to infer latent codes and produces visually appealing hair strands, offering a streamlined alternative to complex multi-view data acquisition setups.  
Given the latent code, PanoHair generates a clean manifold mesh for the hair region in under 5 seconds, along with semantic and orientation maps, marking a significant improvement over existing methods, as demonstrated in our experiments. 
\end{abstract}

%-------------------------------------------------------------------------
\section{Introduction}
\label{sec:intro}

High-fidelity 3D hair modeling is essential for creating realistic digital humans. A hairstyle often comprises tens of thousands of individual strands, which adds significant complexity to the process. 
Despite advancements in high-end capture systems, obtaining high-quality 3D hair models remains challenging, as accurately reconstructing the intricate geometry of hair is still difficult.

The hairstyle generation has traditionally depended on the intricate manual modeling process by expert 3D artists, with strand-based hair geometry being the most commonly adopted approach. 
Earlier methods \cite{chai2013dynamic,chai2012single} involved manipulating hairstyles based on images and strokes as input cues, facilitating faster modeling.
Building on recent progress in learning-based shape reconstruction, neural networks are now capable of generating 3D strand models using explicit point sequences \cite{zhou2018hairnet},  volumetric orientation fields \cite{saito20183d,zhang2019hair,zheng2023hairstep}, and implicit orientation fields \cite{wu2022neuralhdhair,sklyarova2023neural}. 
Recent approaches \cite{wu2022neuralhdhair, sklyarova2023neural, wu2024monohair} first estimate the volume enclosed by the hair region and 3D orientations in the initial stage, followed by a hair-growing procedure in the second stage to generate the final strand-based hairstyle. 
These methods often rely on complex data-processing pipelines that use multi-view images to reconstruct sparse point clouds, estimate camera positions, and perform bust fitting, which provides a shape before learning the fine geometry of the hair region.

% Creating a paired dataset of 3D hair and corresponding real images is challenging. Existing approaches often rely on synthetic data, which, although convenient, offers limited hairstyle diversity and introduces a domain gap between rendered synthetic and real images, negatively affecting model performance. To address this, we aim to leverage generative models trained on large 2D image databases that capture realistic hairstyles. 
Recent 3D-aware generative models \cite{deng2022gram, verma2025semfaceedit, chan2022efficient} have demonstrated impressive results in generating highly realistic novel views. These models also offer straightforward inference by manipulating latent codes, eliminating the need for complex data acquisition and pre-processing pipelines. This latent space manipulation also enables indefinite data generation with novel views.
Nevertheless, to fully harness the potential of these generative models for strand-based hair geometry, it is essential to precisely segment the hair region from the full-head geometry and estimate 3D surface orientations that guide strand growth for realistic hairstyle generation.

In this work, we introduce PanoHair, which leverages knowledge distillation from the pre-trained model, PanoHead \cite{An_2023_CVPR}, to generate the geometry of the complete human head. 
PanoHead serves as the teacher model, guiding PanoHair, the student model, to reconstruct a complete head model.
Through knowledge distillation, PanoHair efficiently learns to generate high-quality signed distance fields (SDF) for the full-head geometry.
Additionally, for each point on the iso-surface in volume, our approach predicts binary semantic information to classify whether it belongs to the hair region, along with a 3D orientation vector. 
% The geometry formed by all points belonging to the hair region defines the 'outer shell' of the hair surface. This geometric boundary helps infer strand occupancy, providing an upper constraint to ensure that hairs do not grow outside the estimated hair region. The predicted orientations at these points guide the strands in following the correct direction.
However, this representation captures only the ``outer shell" and does not fully account for the enclosed hair volume. 
To address this limitation, we propose a simple yet effective method that accurately closes the outer shell to represent the full hair volume. 
This straightforward approach ensures a fast estimation of the fully enclosed hair volume, an essential requirement for the hair-growing stage and realistic strand synthesis. Compared to NeuralHaircut \cite{sklyarova2023neural}, which requires approximately one day for hair geometry estimation, our method performs pivotal inversion given a real image to obtain the latent code, after which the geometry estimation takes just five seconds. The substantial reduction in processing time and the generation of high-quality hairstyles make PanoHair a significant advancement in 3D hair modeling.

\noindent \textbf{Contributions.} In summary, our contributions are as follows:
\textbf{1.} We introduce PanoHair, a generative framework that distills knowledge from PanoHead to model human heads as SDFs while predicting semantic labels and 3D orientations.
    \textbf{2.} We propose a multi-view orientation projection loss to ensure view-consistent 3D orientation learning and mitigate ambiguities from single-view 2D projections.
    \textbf{3.} We employ a fast and robust hair volume extraction method using boolean operations, leveraging the structured and bounded nature of the learned implicit representation.

\section{Related Work}
\label{sec:relw}
We bridge the gap between generative volumetric head modeling and hair strand synthesis. This section reviews existing methods relevant to our approach.

\noindent \textbf{Implicit Representations for Head and Hairs.}
A line of research focuses on parametric head models \cite{li2017learning, giebenhain2023learning}, representing the head as a rigged mesh with a fixed topology.
More recently, implicit representations, such as NeRF \cite{mildenhall2021nerf, muller2022instant} and implicit surfaces \cite{wang2021neus, yariv2020multiview} have emerged for modeling human heads, enabling applications like novel view synthesis and generative face modeling \cite{chan2022efficient, deng2022gram, verma2025semfaceedit}. 
However, like parametric head models, volumetric representations inherently lack support for strand-based hair synthesis.
A common approach involves estimating the hair region's occupancy from the head geometry and then using a data-driven method to achieve realistic hair growth within the bounded volume. 
For hair modeling, earlier methods investigated different structured geometric primitives, such as wisp-based models \cite{watanabe1992trigonal}, parametric surfaces \cite{liang2003enhanced, noble2004modelling}, and cylindrical primitives \cite{choe2005statistical, kim2002interactive, wang2004hair}. 
Strand-based representations for hair are now widely recognized as the standard for high-fidelity hair modeling \cite{piuze2011generalized, shen2023ct2hair, hsu2023sag}. Recent works encode surface strand geometry into latent codes controlling strand length, curvature, and orientation \cite{sklyarova2023neural, sklyarova2024text}.

\noindent \textbf{Generative Modeling of Human Head.}
With significant progress in generating synthetic facial images using Generative Adversarial Networks (GANs) \cite{karras2020analyzing, zhu2020sean} and Denoising Diffusion Probabilistic Models (DDPMs)  \cite{huang2023collaborative, kim2022diffusionclip, ding2023diffusionrig}, recent approaches leverage 3D-aware supervision to produce volumetric face models as view-consistent implicit representations. 
Models trained on large datasets of frontal 2D human faces \cite{deng2022gram, chan2022efficient, verma2025semfaceedit, jiang2022nerffaceediting} are limited to generating geometry and novel view renderings of frontal faces, restricting their ability to estimate the complete geometry of a human head.
This limitation restricts their applicability in generative hair modeling, as a substantial portion of hairstyle geometry remains inadequately captured.
Recent approaches \cite{An_2023_CVPR, zielonka2023instant} overcome this limitation by capturing the full volumetric geometry of the head, paving the way for strand-based hair modeling, a challenge we address in this work.

\begin{figure*}[t]
    \centering
    \includegraphics[width=\linewidth ]{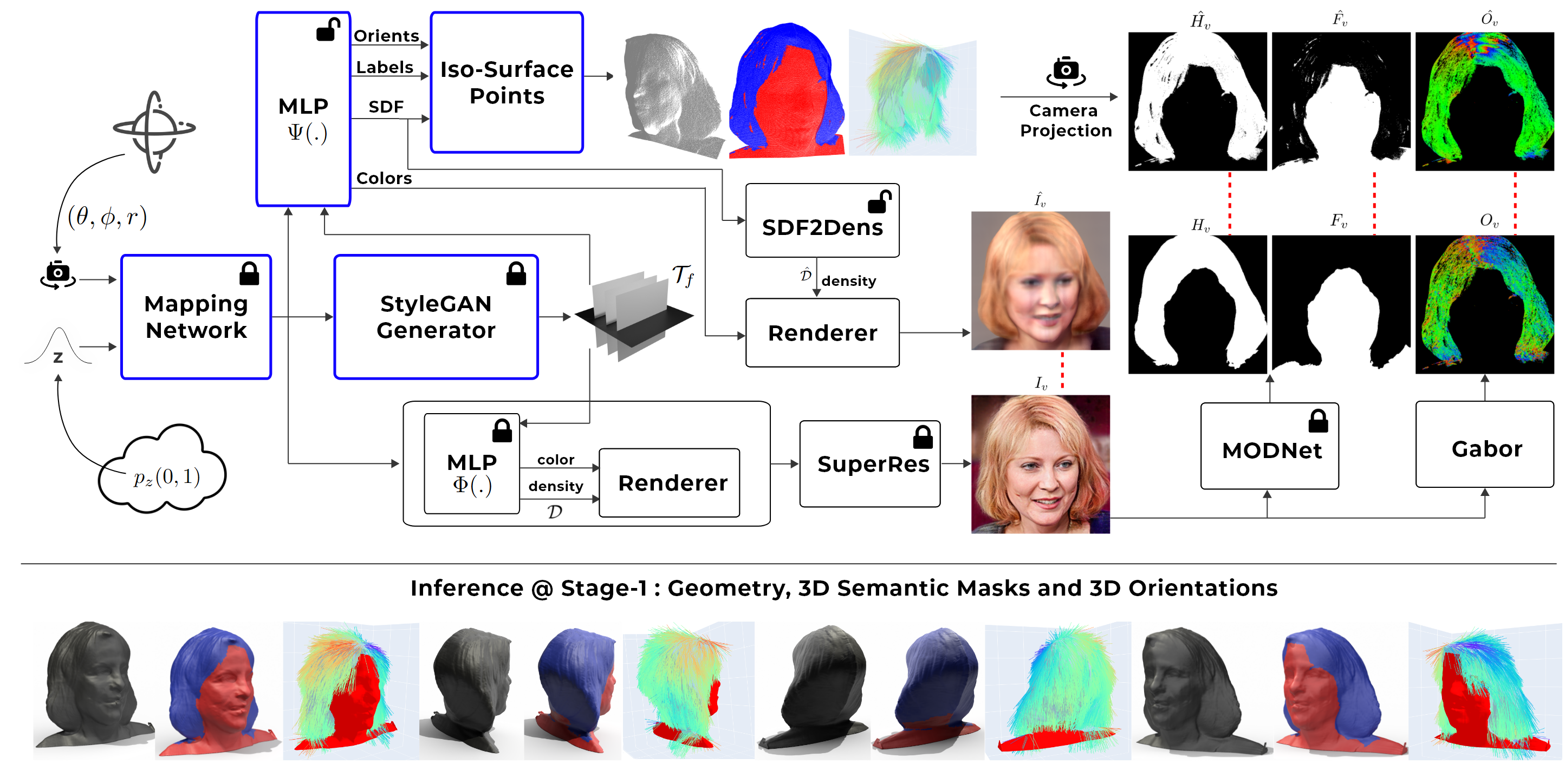}
    \caption{
    \textbf{PanoHair Overview.}
    Given latent code $z$ and camera parameters $C_{\text{cam}} = (\theta,\phi,r)$, we first obtain tri-grid features $\mathcal{T}_f$ using the pre-trained PanoHead \cite{An_2023_CVPR}. A set of MLPs $\Psi(.)$ then predicts semantic labels, 3D orientations, and SDF values. Training is supervised by distilling knowledge from PanoHair image renderings, Gabor orientations, and semantic segmentation. During inference, only the Blue-boxed components are required. The bottom row illustrates PanoHair’s outputs—geometry, semantic labels, and 3D orientations.
    }
    \label{fig:arch}
\end{figure*}
\noindent \textbf{Strand based Hair Reconstruction.}
Image-based hair modeling \cite{nam2019strand, yang2019dynamic, zhang2019hair, luo2013wide, wu2024monohair} has primarily relied on orientation maps estimated from multiple-views \cite{paris2004capture, granlund1978search}. 
However, orientation maps can only supervise the outermost hair surface, resulting in ambiguity when reconstructing the internal hair strand structure. Furthermore, these methods typically demand intricate data preprocessing, including camera localization, sparse point-cloud reconstruction, and bust fitting.
Data-driven methods address this limitation by incorporating hair structure priors. HairStep \cite{zheng2023hairstep} constructs a database of manually annotated orientations to resolve directional ambiguities present in Gabor filter-based orientation estimations. Meanwhile, \cite{zakharov2024human, sklyarova2024text} utilize diffusion priors to enhance the accuracy and fidelity of hair geometry refinement.
NeuralHaircut \cite{sklyarova2023neural} first extracts the hair volume by estimating geometry employing \cite{wang2021neus} and 3D orientation maps, guided by multi-view supervision, and then utilizes diffusion-based priors to generate hair strands within the bounded volume. 

\section{Method}

\noindent \textbf{Overview.} 
Given a latent code, our network learns a signed distance function (SDF) to represent the volumetric head. Hair generation proceeds in three stages: (1) predict the SDF, segmentation labels, and 3D orientations; (2) estimate the hair-specific volume; and (3) grow strand-based hair within this volume.
We leverage knowledge distillation from PanoHead \cite{An_2023_CVPR}, which predicts radiance field densities, guiding our SDF-based surface modeling. Our network also predicts binary segmentation labels on the SDF iso-surface to identify hair regions. Supervision for segmentation and orientation comes from Gabor maps and segmentation masks rendered from PanoHead.
To extract the full hair volume, we fit a FLAME model \cite{li2017learning} and apply boolean operations. The FLAME scalp provides structured strand roots, guided by the predicted 3D orientations.

\subsection{Semantic Surface Extraction with Orientations}
We represent the head as an implicit function modeled by $\Psi(.)$, where 
$\Psi$ takes a 3D point as input and predicts its signed distance, a binary segmentation mask, and 3D orientation. 
We now describe how knowledge distillation from PanoHead \cite{An_2023_CVPR} is used to train PanoHair.

% \subsubsection{Network Architecture}
\noindent \textbf{Network Architecture.}
In Figure \ref{fig:arch}, we illustrate the complete pipeline for training $\Psi(.)$, a key module of Panohair for predicting color $c$, signed distance $s$, orientations $o$, and semantic labels $m$. 
Our proposed network, PanoHair, comprises several sub-modules: Mapping Network, Style-GAN Generator, MLP Network $\Psi(.)$, and SDF2Dens Network. Among these, the Mapping Network and Style-GAN Generator are pre-trained sub-modules adopted from PanoHead \cite{An_2023_CVPR}.
In \cite{An_2023_CVPR}, the MLP Network $\Phi(.)$ decodes tri-grid features into color and density values. Our network $\Psi(.)$ follows a similar architecture; however, in contrast, it predicts signed distances along with orientations and semantic labels essential for hair modeling.
Following recent works \cite{wang2021neus,yariv2020multiview}, we aim to model implicit surfaces with image-based reconstruction loss using differentiable volume rendering. To achieve this, we transform SDF values ($s$) to density values for volume rendering employing the SDF2Dens Module, as illustrated in Figure \ref{fig:arch}. The conversion is defined in Equation \ref{eq:sdf2d}, where $\beta$ is a learnable parameter.
\begin{equation}
    \text{SDF2Dens}_{\beta} (s) = \begin{cases} 
    \frac{1}{2\beta} \text{exp}(\frac{s}{\beta}), & \text{if } s \leq 0 \\
    \frac{1}{\beta}(1-\frac{1}{2}\text{exp}(-\frac{s}{\beta}), & \text{if } s > 0
    \end{cases}
    \label{eq:sdf2d}
\end{equation}

 \begin{figure*}[t]
    \centering
    \includegraphics[width=\linewidth ]{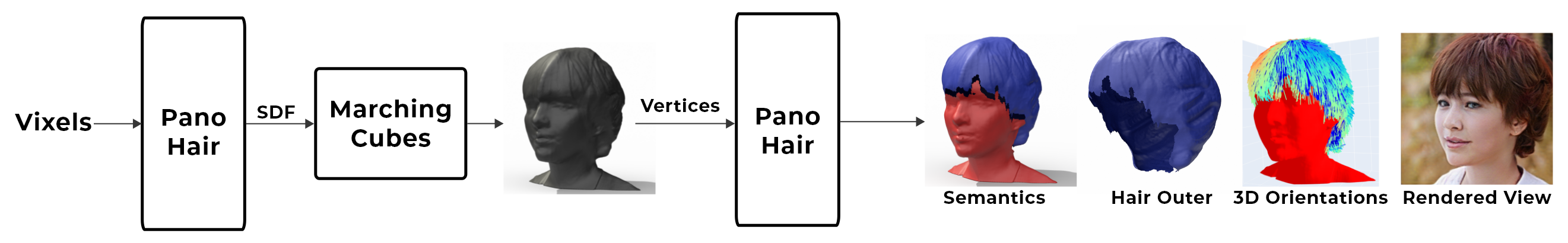}
    \caption{
    Inferencing hair’s outer surface and orientations. A $512 \times 512 \times 512$ volumetric grid is sampled within $[-0.75, 0.75]$ to estimate SDF values. The geometry is extracted using marching cubes, and semantic labels and 3D orientations are computed for the mesh vertices.
    }
    \label{fig:ho}
\end{figure*}

% \begin{figure}[t]
%     \centering
%     \includegraphics[width=0.5\linewidth ]{images/geom.png}
%     \caption{
%     The geometric shapes extracted from implicit representations learned using existing methods: (a) GRAM \cite{deng2022gram}, (b) Tri-Planes \cite{chan2022efficient}, (c) PanoHead \cite{An_2023_CVPR}, and (d) Ours.
%     }
%     \label{fig:geom}
%     \vspace{-0.3cm}
% \end{figure}

% \subsubsection{PanoHead}
\noindent \textbf{Training requirements from PanoHead.} Including details on why PanoHead is preferred over existing methods in the supplementary material, we now outline the training requirements for knowledge distillation using PanoHead.
As shown in Figure \ref{fig:arch}, given a latent code $z \sim p_z(0,1)$ and a desired camera pose $C_{cam}$, the mapping network maps them into an intermediate latent code $w$. The Style-GAN generator takes $w$ as input and produces tri-grid representation features $\mathcal{T}_f$. 
We sample features for a 3D point $\mathbf{x}$ along a casted ray by projecting its coordinates onto each tri-grid and retrieving the corresponding feature vector, $t_{x}$, using trilinear interpolation. 
The MLP Network $\Phi(.)$ takes $t_x$ as input and predicts appearance color and neural radiance density $\mathcal{D}$. 
Finally, volumetric rendering of this radiance field from the given camera views $C_{cam}$ generates the image $I$, with a super-resolved version represented as $I^+$.
We utilize the pre-trained weights from \cite{An_2023_CVPR} and estimate the required components for training PanoHair, as outlined in Equation \ref{eq:ph}. Here, $\mathbf{\text{PH}}(.)$ refers to the PanoHead.
\begin{equation}
    \mathbf{\text{PH}}(z,C_{cam}) \longmapsto (\mathcal{T}_f, I^+, \mathcal{D})
    \label{eq:ph}
\end{equation}
% \subsubsection{Training Strategy}
Let the sampled points along rays cast from camera $C_{cam}$ in the volumetric grid be denoted as $\mathcal{X} \in \mathrm{R}^{N \times 3}$. 
We train $\Psi(.)$ for all $\mathcal{X}$ to predict signed distance from contained surface $\mathcal{S} \in \mathrm{R}^{N \times 1}$, color values $\mathcal{C} \in \mathrm{R}^{N \times 3}$, Semantic labels $\mathcal{M} \in \mathrm{R}^{N \times 1}$, and orientations $\mathcal{O} \in \mathrm{R}^{N \times 3}$, i.e. $\Psi(\mathcal{X}) \longmapsto (\mathcal{S}, \mathcal{C}, \mathcal{M}, \mathcal{O})$.
More precisely, given latent code $w$ and tri-grid features $t_f$ for each point $\mathbf{x} \in \mathcal{X}$ in the volume, we predict signed distance $s \in \mathcal{S}$, color $c \in \mathcal{C}$, binary semantic label $m \in \mathcal{M}$ and orientation $o \in \mathcal{O}$. 

\noindent \textbf{Image Synthesis.}
To facilitate volumetric rendering, we convert the signed distance values  $s \in \mathcal{S}$ into density values $\sigma \in \hat{\mathcal{D}}$  using the SDF2Dens module, as defined in Equation \ref{eq:sdf2d}. 
To synthesize an image $\hat{I}$, we employ a ray-marching approach \cite{mildenhall2021nerf} where rays \( \mathbf{r}\in \mathcal{R} \) are cast from the camera center through each pixel of the image plane into the 3D space. Here, $\mathcal{R}$ represents the set of all rays corresponding to the pixels of an image with resolution \( H \times W \). Along each ray \( \mathbf{r} \), we sample \( K \) discrete points \( \mathbf{x}_r^i \), where \( i \) denotes the index of the sampled point along the ray. For each sampled point \( \mathbf{x}_r^i \), we retrieve its corresponding density \( \sigma_r^i \in \hat{\mathcal{D}} \) and color \( c_r^i \in \mathcal{C} \) from the predicted outputs of \( \Psi(.) \).
The final pixel color \( C(r) \) for a given ray \( \mathbf{r} \) is computed using volume rendering, where each sampled color \( c_r^i \) is accumulated with an opacity-weighted contribution based on the density values \( \sigma_r^i \). This accumulation is performed using the volumetric rendering equation defined in Equation \ref{eq:vrnd}.
\begin{equation}
    \mathcal{C}_r = \sum_{i=1}^K T_r^i \alpha_r^i c_r^i, \quad T_r^i = \prod_{j=1}^{i-1}(1-\alpha_r^j), \quad \alpha_r^i = 1 - e^{-\sigma_r^i\delta_r^i}
    \label{eq:vrnd}
\end{equation}

\noindent \textbf{Surface Points Attributes.} 
We identify the visible surface points with the predicted SDF values at sampled points by detecting the first sign change along each ray. Since the rays are cast from the camera, the first sign change ensures that the detected point lies on the surface and is visible from the given viewpoint.
Let the semantic labels of these identified surface points $x_S \in \mathcal{X}$ be denoted as $m_S \in \mathcal{M}$ and their 3D orientations as $o_S \in \mathcal{O}$.
To obtain the semantic map $\hat{M}_{2D}$, we directly project $m_S$ to image space using the camera matrix of $C_{cam}$.
In contrast, to compute the 2D orientation map $\hat{\mathcal{O}}_{2D}$, we project the 3D orientations $o_S$ into the camera space using Plücker line coordinates \cite{hartley2003multiple}, similar to \cite{wu2024monohair, rosu2022neural}. 
An illustration of $x_S, m_S$, and $o_S$ is shown in Figure \ref{fig:arch} along with their projections onto image plane, namely $\hat{M}_{2D}$ and $\hat{\mathcal{O}}_{2D}$. We include qualitative results of projections in the \textit{supp. material}.
% Figure \ref{fig:oris} shows predicted orientations compared with ground-truth Gabor orientations along with $\hat{M}_{2D}$.

\begin{wrapfigure}{r}{0.5\textwidth} 
    \centering
     \includegraphics[width=0.48\textwidth]{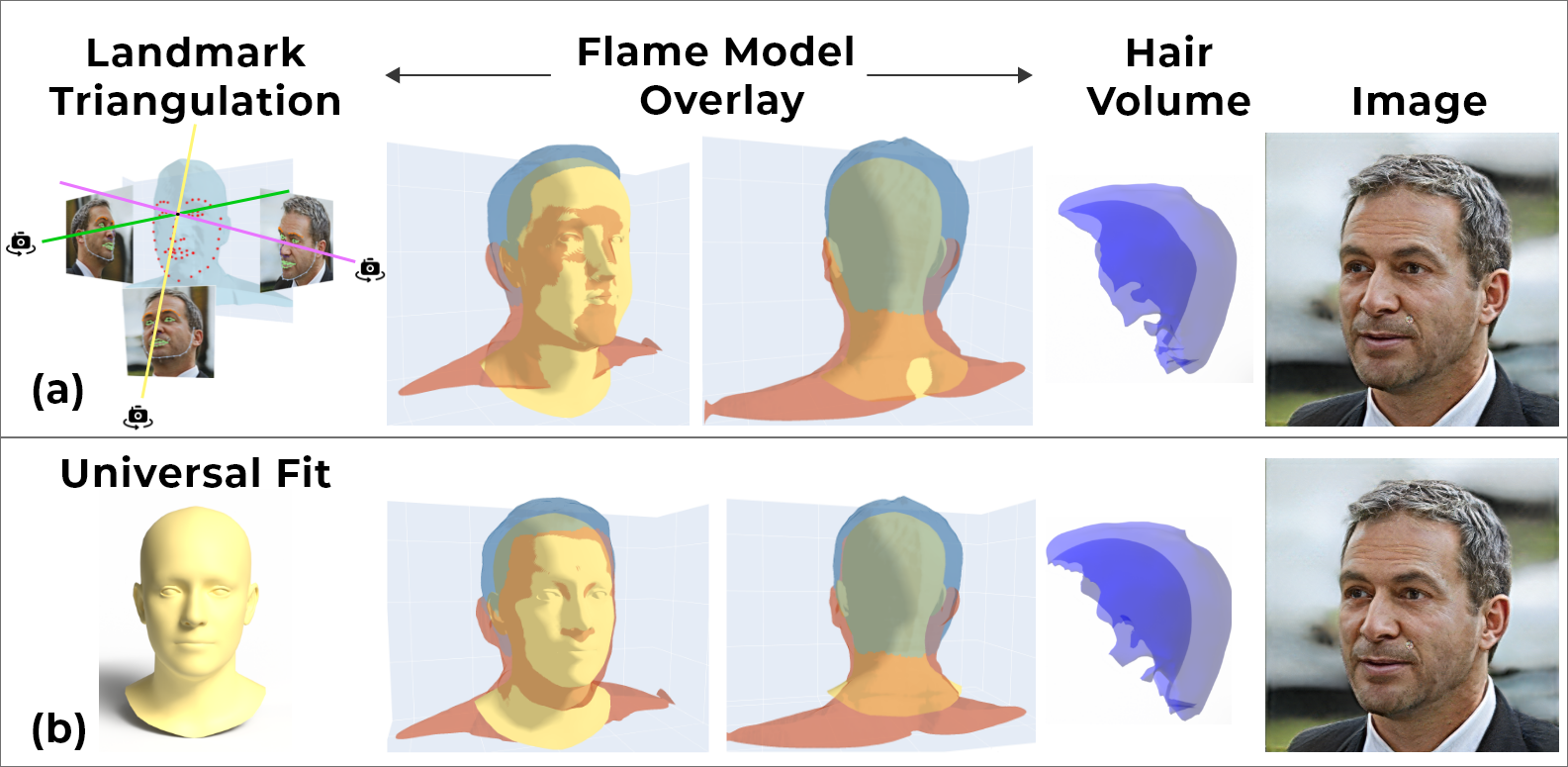}
    \caption{Estimating hair volume. (a) FLAME \cite{li2017learning} model fitted to estimated landmarks. (b) Empirically computed universal fit, leveraging the structured representation of volumetric faces within a unit volume by design during training. Subsequently, we employ Boolean subtraction between the fitted FLAME model and the extracted mesh to compute volume.
    }
    \label{fig:hv}
\end{wrapfigure}
\noindent \textbf{Training Pairs.} 
Our training pairs for learning accurate SDF through volumetric rendering consist of the super-resolved image \( I^+ \) from \cite{An_2023_CVPR} and the synthesized image \( \hat{I} \). To train for semantic masks, we utilize \cite{ke2022modnet} to estimate \( I_M \), which contains hair and facial masks for \( I^+ \), and use them as ground truth for supervision. The training pairs for binary semantic classification of the hair region are \( I_M \) and \( \hat{M}_{2D} \). Furthermore, for \( I^+ \), we estimate a 2D orientation map using a set of Gabor filters to obtain \( I_O \), which serves as the ground truth for the predicted orientation map \( \hat{\mathcal{O}}_{2D} \).

\noindent \textbf{Loss Functions.}
To train PanoHair, we enforce image reconstruction by optimizing a Mean Squared Error (MSE) loss between the super-resolved image \( I^+ \) and the synthesized image \( \hat{I} \). 
For semantic mask prediction, we utilize a Binary Cross-Entropy (BCE) loss between the ground-truth mask \( I_M \) and the predicted mask \( \hat{M}_{2D} \). 
Additionally, to accelerate convergence during the initial training phase, we impose an L1 loss between the density distribution \( \mathcal{D} \) from PanoHead and the estimated densities \( \hat{\mathcal{D}} \), which are derived by transforming the signed distance values predicted by \( \Psi(.) \) using Equation \ref{eq:sdf2d}. 
This auxiliary loss is applied only for the initial training iterations and is gradually reduced to zero afterward using a scheduler $\delta(t)$, which decays to zero after iteration $t$. 
We define our loss function $\mathcal{L}$ in Equation \ref{eq:loss}.

\begin{align}
    \mathcal{L} = 
    &\underbrace{\lambda_{1} \cdot \delta(t) \cdot \| \mathcal{D} - \hat{\mathcal{D}} \|_1}_{\textbf{Density Loss}} + 
    \underbrace{\lambda_{2} \cdot \| I^+ - I \|^2}_{\text{Image Reconstruction Loss}} 
    + \underbrace{\lambda_{3} \cdot \text{BCE}(I_M, \hat{M}_{2D})}_{\textbf{Semantic Loss}} \quad + \quad \mathcal{L}_{\text{orient}}
    \label{eq:loss}
\end{align}

\begin{figure*}[t]
    \centering
    \includegraphics[width=\linewidth ]{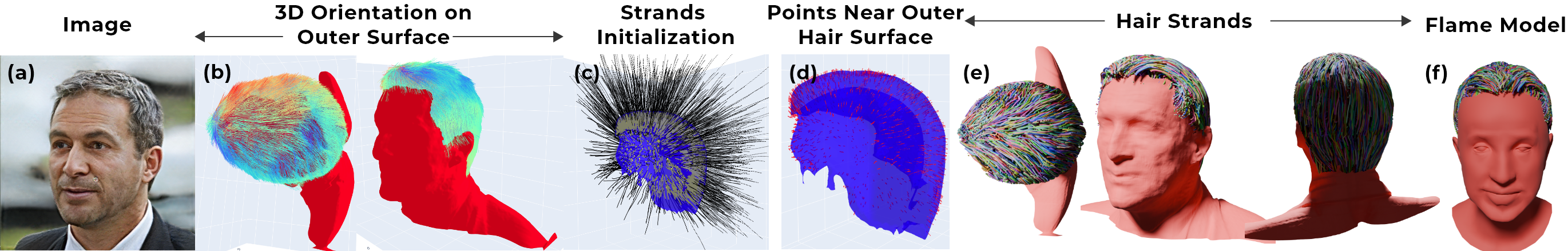}
    \caption{
    Strand-based hair growth. Given a latent code, (a) shows the generated image, and (b) visualizes 3D orientations on the outer hair surface. Strands are initialized using shape textures from \cite{sklyarova2023neural}. Unlike their rasterization-based optimization, we directly optimize boundary strand points (d) to align with predicted orientations (b), producing the final hair (e). (f) shows the FLAME model with strands rooted at the scalp.
    }
    \label{fig:hg}
\end{figure*}
\begin{figure*}[t]
    \centering
    \includegraphics[width=\linewidth ]{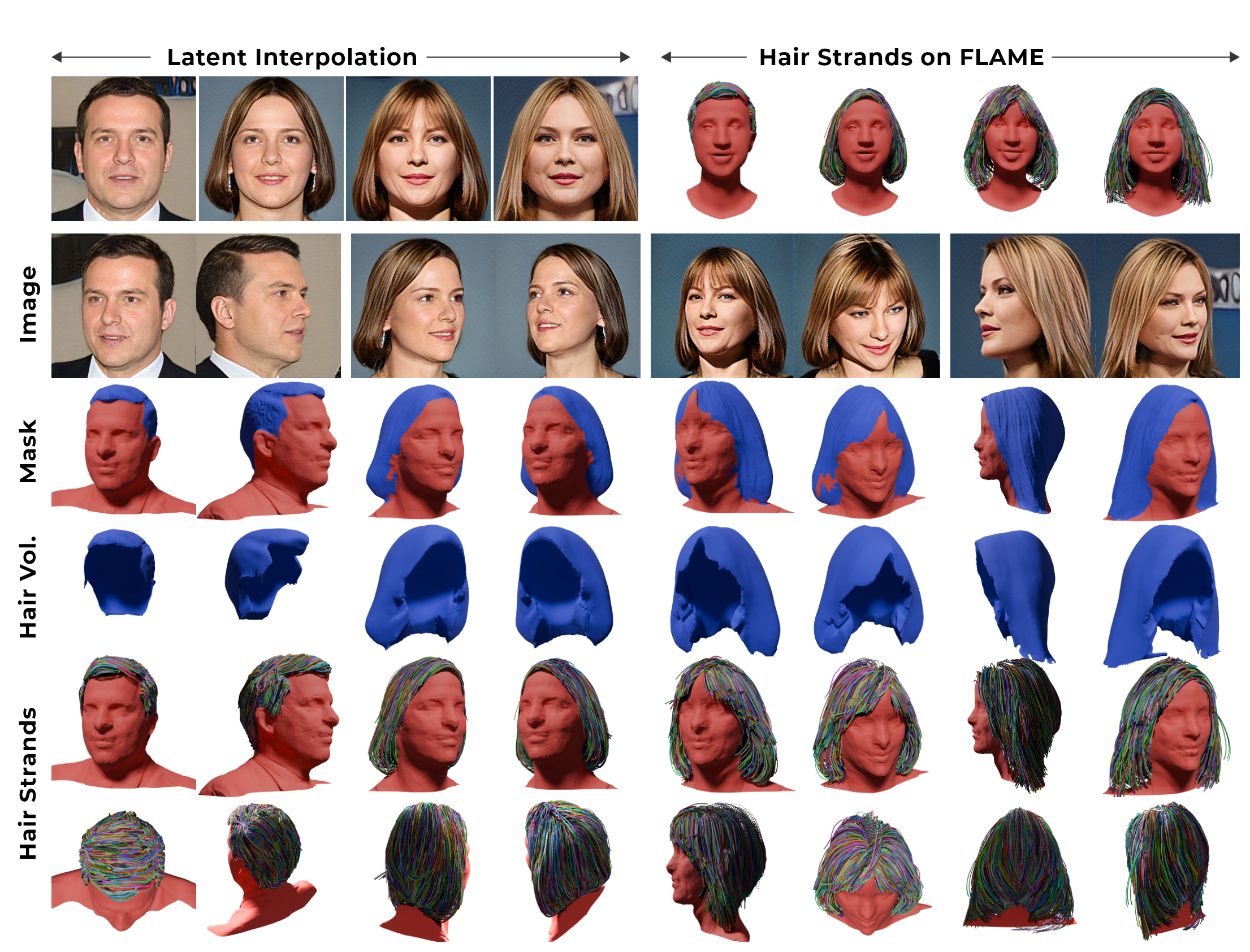}
    \caption{Latent space interpolation between two face codes and hair growth. The second row presents multi-view renderings, followed by semantic masks and computed hair volume in the third and fourth rows. The final rows illustrate the resulting strand-based hair.}
    \label{fig:lint}
\end{figure*}
Here, $\lambda$ represents the weights assigned to each loss term, and $\mathcal{L}_{orient}$ denotes the orientation loss, which we now describe in detail.
Since ground-truth 3D orientations are unavailable for direct supervision, we instead learn to predict them by enforcing losses on their 2D projections, specifically between $\hat{\mathcal{O}_{2D}}$ and the Gabor orientation map $I_O$. However, this projection-based orientation loss is inherently ill-posed and ambiguous, as multiple 3D orientation vectors can result in the same 2D projection on the image plane.
Unlike prior works such as \cite{sklyarova2023neural, rosu2022neural}, which iteratively optimize orientations across multiple views of the same subject, our method is generative, where the identity of the synthesized head is controlled by the latent code $z$. Consequently, the subject changes in each iteration, preventing us from employing such an optimization strategy. Moreover, our predicted orientations must remain consistent across novel views and align with the geometry implicitly represented through the signed distance field (SDF) for each $z$. 
Hence, we impose $\mathcal{L}_\text{orient}$ with two key constraints. First, the predicted orientation vectors must be tangential to the surface. Second, to mitigate the ambiguity introduced by projection, the learning process should enforce multi-view consistency in each iteration for every latent code $z$. Mathematical formulations, training weights, and additional details on $\mathcal{L}_\text{orient}$ are provided in the \textit{supp material}.

\noindent \textbf{Inference.}
During inference for novel-view generation, only the Blue boxes in Figure \ref{fig:arch} are required. SDF-based iso-surfaces yield point cloud geometry suited for gradient propagation via point-based rendering in training. To extract a full head mesh, we sample voxels in a volumetric grid, apply marching cubes \cite{lorensen1998marching} on SDF values, and evaluate semantic masks and orientations at mesh vertices. By selecting points classified as hair, we obtain the hair's outer surface and 3D orientations, as shown in Figure \ref{fig:ho}.

\subsection{Hair Volume Estimation and Growing}
We require a closed, bounded volume where hair strands can reside to facilitate hair growth. While we have estimated the hair’s outer surface, we must also determine the scalp location within the generated head. 
\begin{table*}[t]
\scriptsize
\centering
% \resizebox{0.9\textwidth}{!}{%
    \begin{tabular}{l|c|cc|cc}
    \hline
    \multicolumn{1}{c|}{\textbf{Model}} & \textbf{Input} & \textbf{\begin{tabular}[c]{@{}c@{}}Hair Seg. \\ (IoU)\end{tabular}} & \textbf{\begin{tabular}[c]{@{}c@{}}Orient. Error at \\ Hair Region $(\mathcal{L}_\text{proj})$\end{tabular}} & \textbf{\begin{tabular}[c]{@{}c@{}}Hair Volume Est.\\ time\end{tabular}} & \textbf{\begin{tabular}[c]{@{}c@{}}Strand Reconstruction\\ time\end{tabular}} \\ \hline
    \textbf{NeuralHaircut} \cite{sklyarova2023neural} & Multi-View & 0.815 & 0.391 & 10-12 Hrs & 15-18 Hrs \\
    \textbf{HairStep} \cite{zheng2023hairstep} & Single View & 0.682 & 0.526 & \textbf{$\sim$3 sec} & \textbf{$\sim$5 mins.} \\
    \textbf{PanoHair (Ours)} & Latent code & \textbf{0.846} & \textbf{0.368} & $\sim$5 sec & 4-5 Hrs. \\ \hline
    \end{tabular}%
% }
\vspace{0.2cm}
\caption{We report IoU between the 2D projection of the hair surface and MODNet \cite{ke2022modnet} output as ground truth. Orientation error is measured using $\mathcal{L}_{proj}$ between the projected orientation $\hat{\mathcal{O}}_{2D}$ and high-confidence Gabor orientations. Additionally, we report the total runtime on an NVIDIA RTX-4090 GPU for a single head.
}
\label{tab:sota}
\end{table*}
\noindent \textbf{Hair Volume.}
One approach is to fit a FLAME head model \cite{li2017learning} onto the extracted geometry, which requires identifying known landmark locations on the mesh.
Let $\mathcal{G}$ represent the extracted full head geometry from PanoHair, and let $^iI^+$ be a novel view rendered from $^iC_\text{cam}$. 
We estimate facial landmarks for each view using \cite{bulat2017far}. We use triangulation to compute landmark positions on $\mathcal{G}$ given the known camera parameters. We then align the FLAME model landmarks with the estimated ones through iterative fitting, optimizing shape, expression, and pose parameters to minimize landmark distance \cite{li2017learning}, as illustrated in Figure \ref{fig:hv}(a).
Since the learned implicit representation produces structured geometries $\mathcal{G}$ in terms of pose and bounds, we empirically determine a translation $\mathbf{t_f}$ and scale $\mathbf{s}_f$ for a neutral FLAME model to position the FLAME scalp appropriately, as shown in Figure \ref{fig:hv}(b). We call this fitting Universal Fit, which robustly fits all possible generated heads. While this fit may be less accurate for facial regions, it is sufficient for scalp localization, which is our primary focus. Let the fitted FLAME model geometry be represented as $\mathcal{G}_{F}$
With the approximate locations of both the scalp and the hair's outer surface, we extract a closed, bounded volume for the hair region. To achieve this, we first extrude the hair’s outer surface toward the scalp, ensuring that the extrusion is sufficiently scaled so that the resulting geometry remains entirely within the FLAME model. However, this extrusion process can introduce a non-manifold mesh topology unsuitable for direct Boolean operations.
To address this, we convert both the FLAME head geometry ($\mathcal{G}_F$) and the extruded hair mesh into a volumetric representation and perform a Boolean subtraction, yielding a closed, bounded hair volume $\mathcal{G}_h$.
Figure \ref{fig:hv} qualitatively compares two approaches for obtaining $\mathcal{G}_h$.
% —one involving multi-view optimization and the other a simpler, direct method.

\noindent \textbf{Hair Strand Growing.}
Given the 3D orientations $\mathcal{O}$ at surface points $x_S \in \mathcal{X}$ labeled with hair semantics $\mathcal{M}$, and the hair volume $\mathcal{G}_h$, we adopt a strand-based geometry reconstruction approach similar to \cite{sklyarova2023neural}. Figure \ref{fig:hg} illustrates the optimization steps involved. For more details on strand reconstruction, we refer readers to \cite{sklyarova2023neural}. In contrast to \cite{sklyarova2023neural}, we exclude silhouette and RGB reconstruction losses on multi-view inputs and instead guide strand growth using 3D orientations and the hair volume $\mathcal{G}_h$ predicted by PanoHair.

\begin{figure*}[t]
    \centering
    \includegraphics[width=0.95\linewidth ]{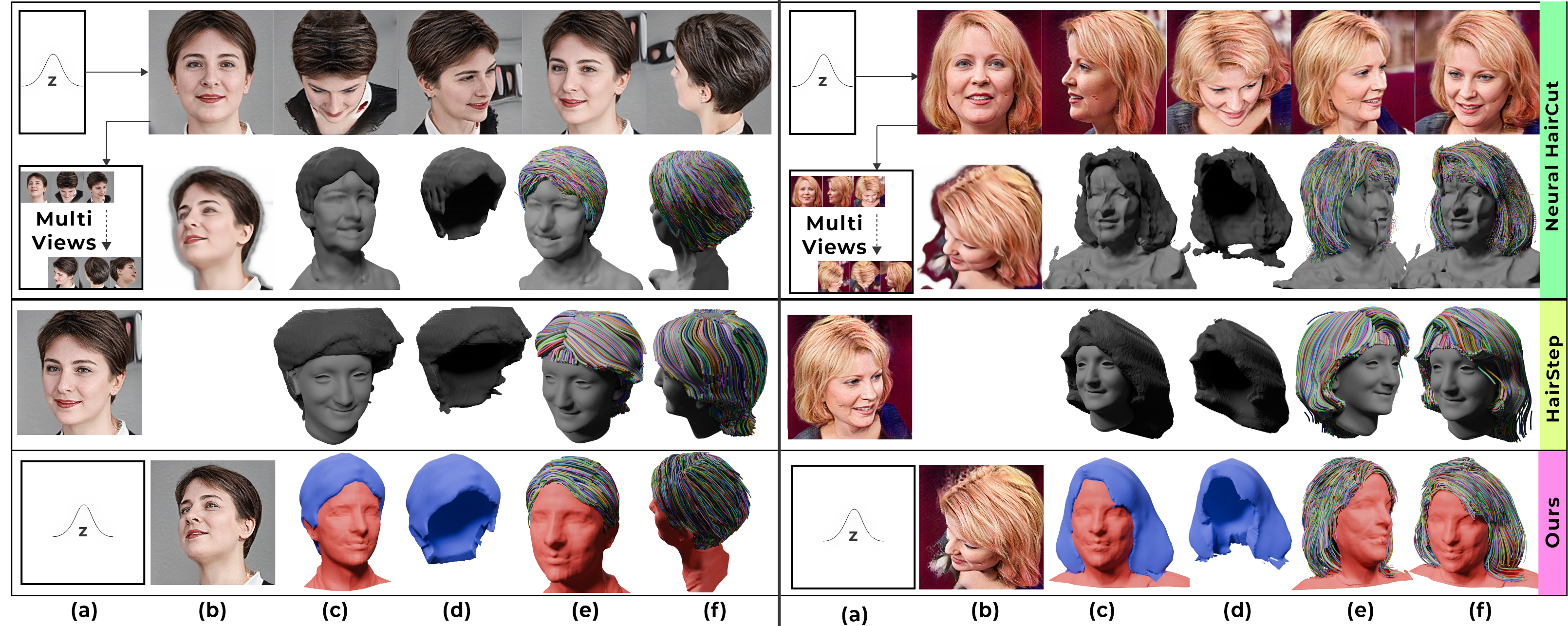}
    \caption{Qualitative comparison of the proposed approach with state-of-the-art methods, NeuralHaircut \cite{sklyarova2023neural} and HairStep \cite{zheng2023hairstep}. NeuralHaircut is trained using multi-view renderings to estimate hair volume and orientations, while HairStep takes a single generated image as input. (a) depicts the input for each method, (b) presents a novel view, (c) and (d) show the full and hair volume meshes, and (e) and (f) illustrate strand-based hair from two views.}
    \label{fig:sota}
    \vspace{-0.2cm}
\end{figure*}

\section{Discussion}
\begin{wrapfigure}{r}{0.5\textwidth} 
    \centering
     \includegraphics[width=0.48\textwidth]{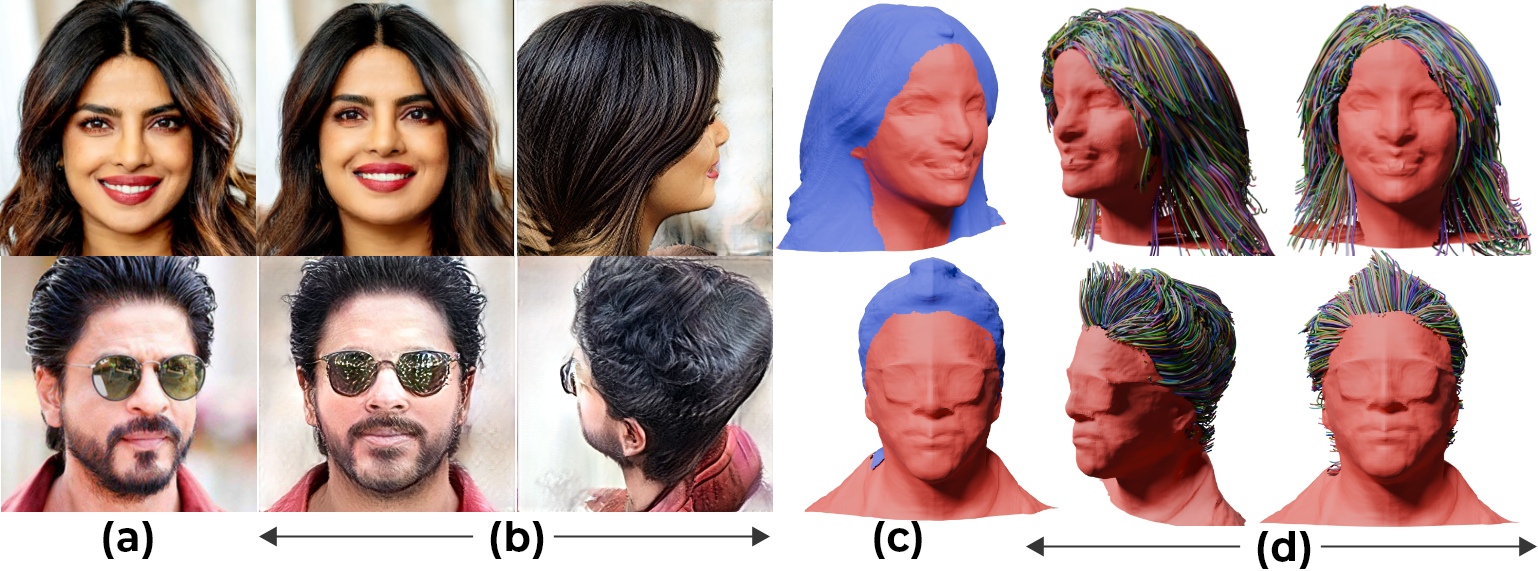}
    \caption{Latent code optimization using Pivotal Inversion \cite{roich2022pivotal} to match the given real image (a). The optimized latent code generates novel views (b), hair masks (c), and 3D orientations. The final strand-based hair reconstruction is shown in (d).}
    \label{fig:inv}
\end{wrapfigure}

\noindent \textbf{Latent Space Exploration and Inversion.}
The generative design of PanoHair ensures 3D orientations consistent with geometry, enabling diverse hairstyle synthesis without relying on limited and hard-to-obtain 3D synthetic data.
Figure \ref{fig:lint} showcases semantic labels and hair volume geometry extracted by our approach. Smooth transitions in the latent code result in consistent orientation maps, enabling smooth transitions between strand-based hairstyles in linearly interpolated latent space.
To generate hairstyles from a single real-world image, we first optimize the latent code $z$ using an inversion process \cite{roich2022pivotal}. With the obtained latent code, the PanoHair framework predicts hair volume and 3D orientations, enabling novel-view synthesis or hair strand generation, as shown in Figure \ref{fig:inv}.

\noindent \textbf{Comparisons.}
While no existing method directly addresses hair growth in a generative volumetric setting for direct comparison, we evaluate our approach against NeuralHaircut \cite{sklyarova2023neural} and HairStep \cite{zheng2023hairstep} under the following assumptions.
NeuralHaircut requires multi-view data to first estimate hair and bust geometry employing $\cite{wang2021neus}$, while HairStep operates on a single image for input.
We randomly sample five latent codes and generate 40 novel views for each, storing camera parameters, Gabor orientations, and masks required by \cite{sklyarova2023neural}. Of these, 35 views are used for training, while five per code are reserved for evaluation. For HairStep, a single frontal view from the 40 training images serves as input, with the same evaluation data. In contrast, our approach directly utilizes the latent code $z$, which generates the corresponding renderings. The data comprises five different heads, totaling 200 multi-view images, with 25 instances allocated for validation.
Table \ref{tab:sota} presents the quantitative results of different methods and their runtime. Hair segmentation is reported using IoU on validation set camera views, while orientation error is evaluated by computing $\mathcal{L}_\text{orient}$. Our method achieves superior performance in both hair-region estimation and orientation accuracy. Regarding runtime, \cite{sklyarova2023neural} requires per-head training with multi-view inputs, taking approximately 12 hours, whereas our approach completes in around 5 seconds. HairStep is faster but produces poor and inconsistent results across views.

\section{Conclusion}
In this study, we present PanoHair, a 3D-aware generative framework that capitalizes on the capabilities of tri-grid representations for full-head modeling. Our methodology represents geometry as implicit signed distance functions (SDFs) while concurrently predicting semantic labels and view-consistent 3D orientations on iso-surfaces. The inference process for estimating hair volume and orientation is notably efficient, requiring approximately five seconds. The rapid and generative characteristics of PanoHair facilitate the synthesis of a wide variety of hairstyles, and we demonstrate that our predicted three-dimensional orientations significantly enhance the generation of detailed strand-based hair structures. We believe this work will excite many researchers to enhance the framework to perform hair strand synthesis for full-head models. 

\vspace{0.5 cm}
\noindent \textbf{Acknowledgements.} We would like to acknowledge the support of the Jibaben Patel Chair in AI for this work.

\bibliography{main}
% {
%     % \small
%     % \bibliographystyle{ieeenat_fullname}
%     \bibliography{main}
% }

\clearpage
\section*{Appendix}
\section{PanoHead}
\begin{figure}[t]
    \centering
    \includegraphics[width=0.7\linewidth ]{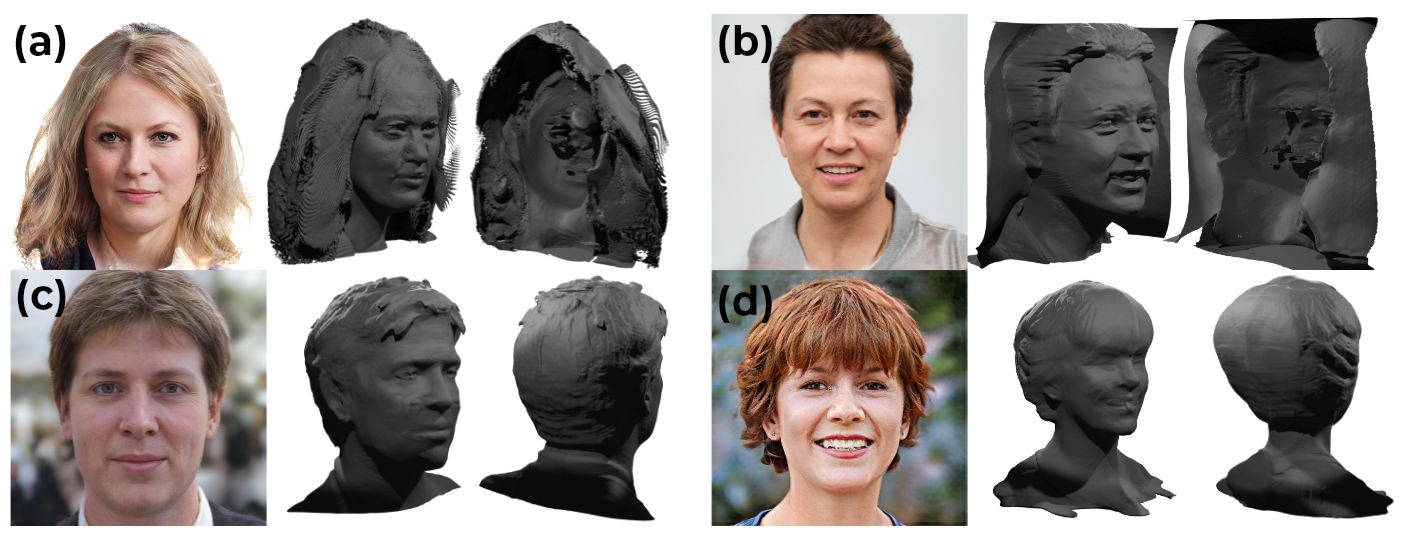}
    \caption{
    The geometric shapes extracted from implicit representations learned using existing methods: (a) GRAM \cite{deng2022gram}, (b) Tri-Planes \cite{chan2022efficient}, (c) PanoHead \cite{An_2023_CVPR}, and (d) Ours.
    }
    \label{fig:geom}
    \vspace{-0.3cm}
\end{figure}

\noindent \textbf{Why PanoHead Over Existing Approaches?}
Various works have explored implicit volumetric representations of the human head. GRAM \cite{deng2022gram} utilizes structured manifolds, while \cite{chan2022efficient} employs Tri-Plane representations. PanoHead \cite{An_2023_CVPR} further advances this by leveraging Tri-Grid feature representations. 
As illustrated in Figure \ref{fig:geom}, methods trained solely on frontal views struggle to accurately capture the complete 360$^{\circ}$ geometry of the human head, posing a challenge for precise hair volume estimation. Figures \ref{fig:geom}(a) and \ref{fig:geom}(b) depict the geometric reconstructions obtained by \cite{deng2022gram} and \cite{chan2022efficient}, respectively. PanoHead \cite{An_2023_CVPR} overcomes this limitation by utilizing tri-grid feature representation $\mathcal{T}_f$, enabling the reconstruction of the full 360$^{\circ}$ head geometry, as shown in Figure \ref{fig:geom}(c). 
However, PanoHead is trained to predict the color and density distribution within a volumetric representation by design, enabling view-consistent neural radiance field rendering. 
This radiance-based implicit representation, modeled as a volumetric density field, is optimized for rendering but suboptimal for surface extraction. Therefore, we propose to distill knowledge from PanoHead by training 
$\Psi(.)$ to directly predict signed distances. 
By leveraging the signed distance function (SDF), the surface can be naturally extracted as the zero-level set or iso-surface, providing a more precise and well-defined geometry.
Moreover, we extend $\Psi(.)$ to also predict semantic masks and 3D orientations for every point on the iso-surface of the predicted SDF, enabling a more structured understanding of hair geometry. This comprehensive approach ensures that PanoHair captures both the geometry and semantics necessary for high-fidelity hair modeling.

The ability of tri-grid representations to capture full-head geometry, as demonstrated by PanoHead \cite{An_2023_CVPR}, forms the foundation of PanoHair. Figure \ref{fig:geom} qualitatively compares head geometries obtained from various 3D-aware neural representations. Generative Manifolds \cite{deng2022gram} and Tri-planes \cite{chan2022efficient} primarily reconstruct frontal geometries (Figure \ref{fig:geom}(a), (b)). In contrast, we leverage knowledge distillation from PanoHead to learn full-head SDFs (Figure \ref{fig:geom}(d)), whereas PanoHead represents faces as density distributions (Figure \ref{fig:geom}(c)). 

\section{Loss Functions}
In the main paper, we define our loss function $\mathcal{L}$ as presented in Equation \ref{eq:loss}
\begin{align}
    \mathcal{L} = 
    &\underbrace{\lambda_{1} \cdot \delta(t) \cdot \| \mathcal{D} - \hat{\mathcal{D}} \|_1}_{\textbf{Density Loss}} + 
    \underbrace{\lambda_{2} \cdot \| I^+ - I \|^2}_{\text{Image Reconstruction Loss}} \notag \\
    &+ \underbrace{\lambda_{3} \cdot \text{BCE}(I_M, \hat{M}_{2D})}_{\textbf{Semantic Loss}} \quad + \quad \mathcal{L}_{\text{orient}}
    \label{eq:suploss}
\end{align}
Here, $\lambda$ represents the weights assigned to each loss term, and $\mathcal{L}_{orient}$ denotes the orientation loss, which we now describe in detail.
\subsection{Tangential Loss.} 
Since the SDF $\mathcal{S}$ encodes signed distances, its gradient at the surface points $x_S$ points in the direction of the outward surface normal. We impose tangential loss for orientations $o_S$ predicted at surface points as defined in equation \ref{eq:tanloss}.
\begin{equation}
   \mathcal{L}_{\text{tan}} = \frac{1}{N}\sum_{i=1}^N | o^i_S \cdot n_S^i | \quad \text{where} \quad n_S = \frac{\nabla \mathcal{S}}{||\nabla \mathcal{S}||}
   \label{eq:tanloss}
\end{equation}

\subsection{Multi-view Orientation Projection Loss.}
We generate \( k \) multi-view outputs from PanoHead using cameras \( {^iC_\text{cam}} \) and render the corresponding super-resolved images \( I^+ \) along with their Gabor orientation maps \( I_O \). For clarity, we adopt the notation \( {^iX} \), where the superscript \( i \) on the top left of a variable \( X \) indicates its association with the \( i \)-th view.  
To ensure consistency across views, we synthesize these \( k \) views so that each image overlaps by more than 70\% with the first view. For surface points \( {^0x_S} \) identified in the first view, we project them into all \( k \) camera views, including the first view itself, and infer their ground-truth Gabor orientations from \( {^iI_O} \). 
Gabor orientation maps represent local image structures by capturing dominant edge directions. However, they exhibit directional ambiguity because an orientation vector $\theta$ and its opposite direction $\theta + \pi$ produce the same response. This means that for a predicted orientation $^0\hat{\mathcal{O}}_{2D}$  its true counterpart $I_O$ can be any of the $I_O, I_O+\pi,$ and $I_O-\pi$.
Hence, for projected predicted orientations $^0\hat{\mathcal{O}}_{2D}$ in the first view, we impose projection loss as in Equation \ref{eq:orloss}.
This enables a multi-view consistent optimization of the orientation vector. 
\begin{align}
    \mathcal{L}_{\text{proj}} = 
    &\sum_{i=0}^{k-1} \min \Big( \left| {^0\hat{\mathcal{O}}_{2D}} - {^iI_O} \right|,  
    \left| {^0\hat{\mathcal{O}}_{2D}} - {^iI_O} - \pi \right|,  \notag\\
    &\quad \left| {^0\hat{\mathcal{O}}_{2D}} - {^iI_O} + \pi \right| \Big)
    \label{eq:orloss}
\end{align}

\begin{figure}[t]
    \centering
    \includegraphics[width=0.7\linewidth ]{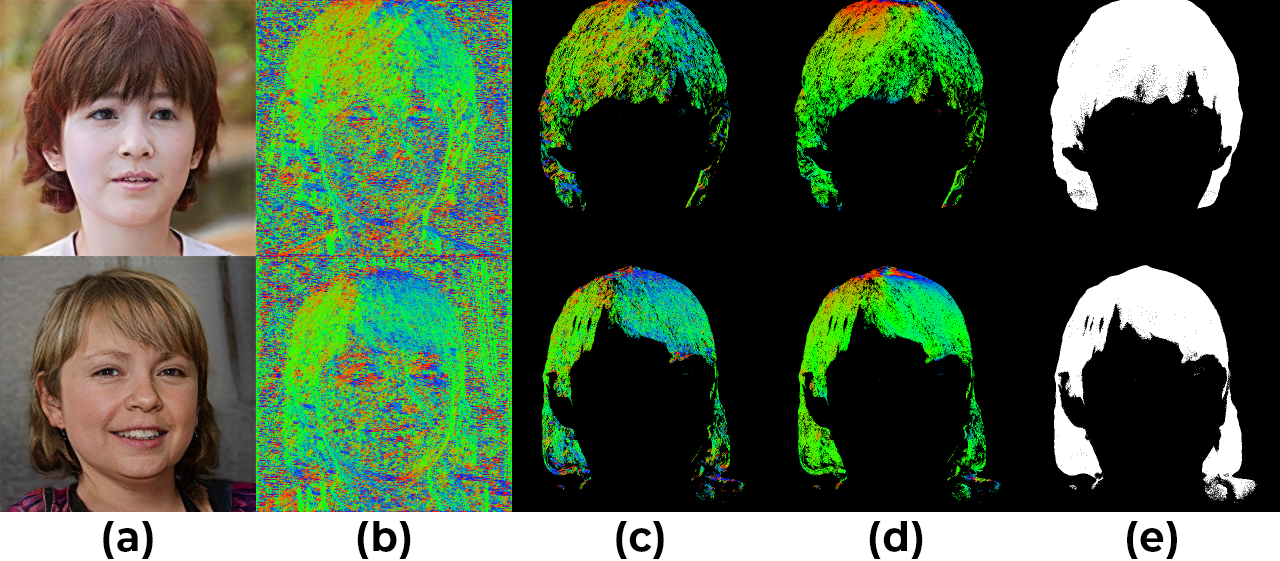}
    \caption{For the image (a) generated from a latent code, (b) shows estimated Gabor orientations, while (c) highlights high-confidence ($>0.2$) orientations in the hair region. Predicted orientations for hair region points are shown in (d), and screen-projected locations of pixels belonging to hair semantics in (e).}
    \label{fig:oris}
    \vspace{-0.2cm}
\end{figure}

\begin{table}[t]
\centering
% \resizebox{\textwidth}{!}{%
\begin{tabular}{c|cc|cc}
\hline
Loss & $L_\text{tan}$ & $L_\text{proj}$ & $L_\text{tan}$ & $L_\text{proj} \;\;(rad)$ \\ \hline
Multi-view & \cmark & \cmark & 0.16 & 0.31 \\ \hline
Multi-view & \xmark & \cmark & 0.20 & 0.34 \\ \hline
Single-view & \cmark & \cmark & 0.18 & 0.42 \\ \hline
Single-view & \xmark & \cmark & 0.26 & 0.48 \\ \hline
\end{tabular}%
% }
\vspace{0.2cm}
\caption{Impact of training PanoHair with and without multiple-views in consideration }
\label{tab:ablloss}
\end{table}

Finally, the total orientation loss is defined as in equation \ref{eq:lorient}.
\begin{equation}
    \mathcal{L}_\text{orient} = \lambda_4\mathcal{L}_\text{tan} + \lambda_5 \mathcal{L}_\text{proj}
    \label{eq:lorient}
\end{equation}
We train PanoHair for 1 million iterations, sampling latent codes $z$ from a normal distribution $p_z(0,1)$, using the loss functions defined in Equation \ref{eq:suploss}. For multi-view projection loss, we randomly sample $k=3$ multiple overlapping views. The weighting parameters are set as follows: $\lambda_1 = 1e-4$ (decaying to 0 after 20K iterations), $\lambda_2 = 1$, $\lambda_3 = 10$, and $\lambda_4 = \lambda_5 = 10$. Training is conducted on an NVIDIA RTX-4090 GPU and takes approximately three days.
Figure \ref{fig:oris} shows predicted orientations compared with ground-truth Gabor orientations along with $\hat{M}_{2D}$.
In Table \ref{tab:ablloss}, we present an ablation study analyzing the impact of training PanoHair with a single-view orientation projection loss versus a multi-view projection loss. For each setting, we also evaluate the effect of omitting the tangential loss. The results show that the model trained with both the tangential and multi-view projection losses achieves the best performance, yielding the lowest angular error when projecting predicted orientations onto the ground truth. Moreover, the predicted orientations remain more tangential to the surface of the hair mesh, which is desirable. The evaluation is conducted by randomly sampling 100 latent codes.

\end{document}